\documentclass[conference]{IEEEtran}
\IEEEoverridecommandlockouts
\usepackage{cite}
\usepackage{amsmath,amssymb,amsfonts}
\usepackage{algorithmic}
\usepackage{graphicx}
\usepackage{subcaption}
\usepackage{textcomp}
\usepackage{xcolor}
\usepackage{siunitx}
\usepackage{soul}
\def\BibTeX{{\rm B\kern-.05em{\sc i\kern-.025em b}\kern-.08em
    T\kern-.1667em\lower.7ex\hbox{E}\kern-.125emX}}
\begin{document}

\title{Automatic Testing and Validation of Level of Detail Reductions Through Supervised Learning}

\author{\IEEEauthorblockN{}
\IEEEauthorblockA{Matilda Tamm$^{1\dagger}$, Olivia Shamon$^{2\dagger}$, Hector Anadon Leon$^3$, Konrad Tollmar$^3$, Linus Gisslén$^3$\\
\textit{$^1$Uppsala University}\\
\textit{$^2$Linköping University}\\
\textit{$^3$SEED - Electronic Arts (EA)}\\
mata1639@student.uu.se, olish585@student.liu.se, \{hleon, ktollmar, lgisslen\}@ea.com}
}

\maketitle

\begin{abstract}
Modern video games are rapidly growing in size and scale, and to create rich and interesting environments, a large amount of content is needed. As a consequence, often several thousands of detailed 3D assets are used to create a single scene. As each asset's polygon mesh can contain millions of polygons, the number of polygons that need to be drawn every frame may exceed several billions. Therefore, the computational resources often limit how many detailed objects that can be displayed in a scene.
To push this limit and to optimize performance one can reduce the polygon count of the assets when possible. Basically, the idea is that an object at farther distance from the capturing camera, consequently with relatively smaller screen size, its polygon count may be reduced without affecting the perceived quality. Level of Detail (LOD) refers to the complexity level of a 3D model representation. The process of removing complexity is often called LOD reduction and can be done automatically with an algorithm or by hand by artists. 
However, this process may lead to deterioration of the visual quality if the different LODs differ significantly, or if LOD reduction transition is not seamless.
Today the validation of these results is mainly done manually requiring an expert to visually inspect the results. However, this process is slow, mundane, and therefore prone to error.
Herein we propose a method to automate this process based on the use of deep convolutional networks. We report promising results and envision that this method can be used to automate the process of LOD reduction testing and validation.

\end{abstract}

\let\thefootnote\relax\footnotetext{$^\dagger$ Equal contribution}

\section{Introduction}
\begin{figure*}[!h]
    \centering
    \includegraphics[width=0.80\textwidth]{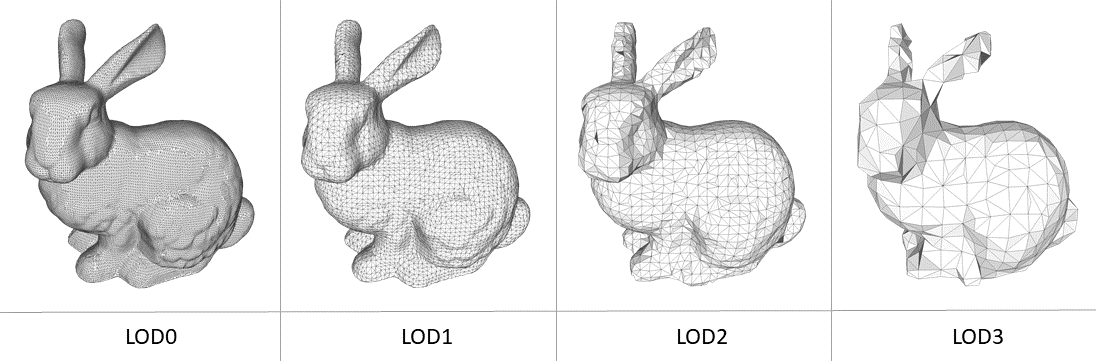}
    \caption{Examples of different LOD levels. When changing the LOD level, there are two distinct effects: first, the surface contacts points are reduced and shifted; second, the silhouette is changed. They usually have a high impact on the perceived quality of the LOD. Asset from \cite{turk2005stanford}.}
    \label{fig:lod_concept}
\end{figure*}

Level of Detail (LOD) refers to the complexity of a 3D model representation. In a 3D video game, it is used to refer to an asset's LOD which is related to an asset's polygon count. An asset can be characters, objects, maps, audio, etc. and are created by game artists. A polygon is a closed figure, usually a triangle, and they are used to create 3D assets in video games. The higher the polygon count is, the higher the resolution of the asset and the higher the LOD, see an example in Figure \ref{fig:lod_concept}. When an artist creates an asset in full resolution/polygon count, they often create several reduced versions (low-polygon count) of the same object so that several seamless transitions can happen when moving away from the object.

The process of transition to less mesh complexity is often called LOD reduction. The reason why LOD reductions are needed is to be able to optimize the performance of the game when rendering images \cite{lod3dgraphics}, which is mainly limited by the computational resources. A game developer needs to consider other things than graphics that also use the CPU/GPU, for example, larger worlds or more levels, so there is a trade-off. LOD0 represents the highest LOD, i.e. the original mesh. The sequential values (LOD1, LOD2, LOD3, etc.) represent lower levels where the highest number represents the lower level of detail. A lower LOD level replaces a higher one as the observing camera moves away from a view and objects become relatively smaller. When an object becomes more distant, it naturally does not have to use as high of a polygon count as fewer details are needed for the player to perceive the asset for what it is and not perceive it as poorly rendered. When assets are shown with a higher LOD level and thus lower polygon count, this decreases the time and computing power needed to render the asset. This leads to a better gaming experience for the player as the game graphics will be smoother and with less lag while conserving high visual quality.

Different metrics can be used to determine which LOD is used, and the most common are either to use the distance between asset and player position or asset screen size-ratio. The latter is less sensitive to screen resolutions and field-of-view settings but the concept is similar. These values need to be set for when each transition between two LOD levels should occur. By LOD validation we refer to the process of validating whether these transition points are correctly chosen so there is no/minimal perceptual visual change on the asset's appearance. A properly chosen point should be optimized so that the higher LOD level should be used as sparsely as possible since these require more computations to render. This means that when a player is zooming out, the LOD reduction should happen as early as possible while preserving the visual experience - the transition should go unnoticed or at least not draw the player's attention. The opposite should happen when the player is zooming in, i.e. the transition to lower LOD levels should happen as late as possible. 

Setting these transition points can be done automatically, but without validation, it will be unclear if it is set at the optimal distance. The alternative is to have an artist do this manually by finding the right spot to shift. Yet, doing it manually is both expensive, mundane and time consuming. In this case, a person needs to study the asset from different angles and in different lighting, zooming in and out between LOD levels, and determine how good the LOD transition is, for example, on a scale between 1-5. This is why an automatic way of validating the LOD transitions was explored in this project.

Switching between LODs can lead to noticeable visual effects in a few ways. One is that if the change in level of quality is significant, it will be noticeable to the human eye. Furthermore, if the shape of the polygon mesh changes, especially the silhouette, the transition undergoes something usually referred as popping. Popping refers to an undesirable visual effect that occurs when the transition of a 3D object to a different pre-calculated level of detail (LOD) is abrupt and obvious to the viewer. Techniques like geomorphing and LOD blending can reduce visual popping significantly by making the transitions more gradual. However, it will not solve all the problems at hand, and the rendering time might not even be reduced \cite{schwarz2006gpu}. 

To automate this process and resolve some of the above mentioned problems, we propose the use of convolutional neural networks (CNN) to classify whether a LOD transition is occurring and to quantify the quality of that transition. To our knowledge, there is currently no efficient way of validating generated LODs (e.g. from artists, other automated methods, etc.) automatically and doing it manually is both expensive and time consuming. We have identified a few use cases that this method can be applied:
\begin{itemize}
\item A method of automatically detecting low quality transitions either as a real time tool for artists, or batch validation of large datasets.
\item A method to determine the least polygon count needed to create a good transition from a given automated solution to get to a desired level of quality. 
\item For quick feedback on what automated solution to use (e.g. compare Houdini, Maya, in-house tool, etc.)
\item A method to give feedback to an automated LOD reduction tool where different parameters can be changed until the transition is least visible. Requires a feedback loop with the LOD reduction tool.
\item A method to optimize individual or scene LOD transition distances to automatically increase performance.
\item A method to discriminate generated LOD reduction to be used in an adversarial fashion together with a learning method to generate different LOD meshes.
\end{itemize}

\section{Previous Work}

In this paper, we focus on the LOD reduction used in games and game production, where the dimensionality of the stored data is reduced for improving rendering time and performance. This proposed approach is an effort to automate the process of testing games at scale, in a similar vein as \cite{bergdahl2020augmenting, holmgaard2018automated, gordillo2021improving} where machine learning is a part of the testing and validation of the game. Evaluating the quality of LOD reduction is complex as we want to perceive the same information from assets that have reduced quality from the reference.
For that reason, LOD reduction evaluation can also be considered as an image quality comparison problem.
These methods are classified into full-reference, reduced-reference, and no-reference algorithms. 
For this paper, only the first applies as the reference asset with the highest quality is always available.
Chandler \cite{chandler2013seven} provides an extensive survey including the other methods.

Image comparison methods can provide different kinds of information. 
Some approaches can yield visual similarity maps to show the per-pixel spatial differences of the images \cite{andersson2020flip, wolski2018dataset}.
On the other hand, other approaches indicate a global score on the similarity between the images \cite{amirshahi2016image, bosse2017deep, wang2003multiscale, zhang2014vsi, mantiuk2011hdr}.

Visual similarity metrics can give more information about where the issues in lower LOD can occur; however, it defeats the purpose of an automatic approach.
One of the traditional approaches, SSIM \cite{wang2004image} produces a per-pixel similarity map based on the degradation of structural information,
 however, it can lead to abnormal and unnatural results \cite{nilsson2020understanding} which makes it an unreliable method.
Using traditional image processing tools, Andersson et al. \cite{andersson2020flip} produce a map that highlights the difference perceived by humans between two images.
This method uses two parallel pipelines: a color pipeline that applies spatial filters based on the human visual contrast and a feature pipeline where edges are localized. Per-pixel comparison is performed for each pipeline between the reference and test image. As the last step, the features are combined for obtaining the final dissimilarity map.
Wolski et al. \cite{wolski2018dataset} use a data-driven approach in order to detect image distortions between pairs. They train a fully convolutional siamese neural network where one branch encodes the difference between the pair and the other encodes the reference. Features are then concatenated and then deconvolved to reconstruct the visibility map. For training this network, they collect a synthetic dataset of reference and distorted image pairs together with user annotations localizing distortions of different kinds.

This work focuses on getting a global similarity value to indicate LOD transition quality instead of a visibility map.
Wolski et al. \cite{wolski2019selecting} select the optimal texture resolution that finds a compromise between resolution and visual quality. They use a similar approach that \cite{wolski2018dataset}, but they explore different pooling strategies to obtain an image metric for detecting visible distortions.
Our approach can be used to assess more than just texture resolution but also to evaluate different LOD transition qualities between different polygon count meshes.
In the quality metrics realm, we can also find other methods that use neural-based approaches.
Bosse et al. \cite{bosse2017deep} extract features from the distorted and reference images using a siamese CNN. Then, the feature vectors are fused, and a patchwise quality estimate is regressed and pooled for the final global quality estimation.
Amirshahi et al. \cite{amirshahi2016image}. Use different levels of the AlexNet \cite{krizhevsky2012imagenet} architecture to compare features using a histogram-based quality metric. Its mean value corresponds to the final score.
PieAPP \cite{prashnani2018pieapp} predicts the human preference between two images over a reference image using a shared-weight feature extraction network and a Bradley-Terry \cite{bradley1952rank} estimating function to model the user preference.
LPIPS \cite{zhang2018unreasonable} is a perceptual similarity metric between image pairs by calculating feature distance in different networks \cite{iandola2016squeezenet, krizhevsky2012imagenet, simonyan2014very} and different architecture stages.

We can also find quality metric applications related to games for selecting texture resolution \cite{wolski2019selecting, lodTexture3DMaps} and for evaluating textured 3D objects.
Guo et al. \cite{guo2016subjective} evaluates textured mesh quality by linearly combining mesh quality and traditional texture quality metrics. This method is problematic as it requires balancing these two metrics that do not scale to different kinds of assets.
Pan et al. \cite{pan2005quality} study the quality reduction when decrementing the texture resolution and the number of vertices.
Finally, although it is not in the scope of the paper, there are methods that automatically simplificate 3D assets. Hasselgren et al. \cite{hasselgren2021appearance} optimize a latent representation consisting of a triangle mesh and a set of textures using an L1 loss as a visual similarity metric to optimize 3D model simplification instead of using a perceptual loss.

None of the above approaches evaluate the perceptual quality of transitions of states between a pair of images. All of them assume perfect alignment between the reference and target images.
Our approach, similar to \cite{ling2020using}, evaluates the feature extraction from different networks \cite{ma2018shufflenet, he2016deep, simonyan2014very, sandler2018mobilenetv2} to assess LOD transition quality. The pair of images that are being compared are not perfectly aligned as our evaluation targets the detection of a transition, a quality shift that happens when the asset is displaced with respect to the camera view at one specific point.

\section{Method}
\label{sec:method}
We studied two approaches for two different use-cases, herein referred to as binary classification and multi-class classification. 

\begin{figure}[!h]
    \centering
    \includegraphics[width=0.50\textwidth]{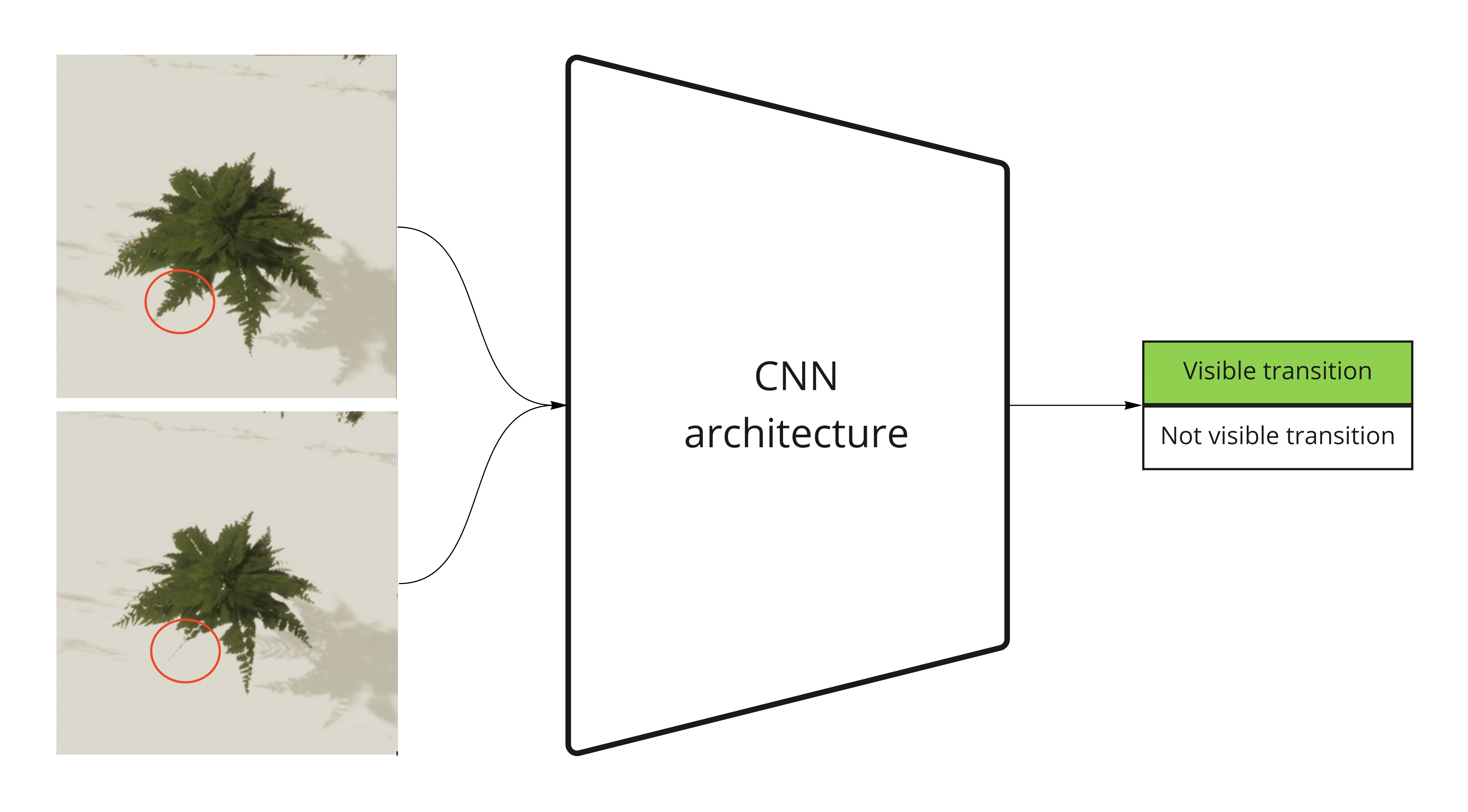}
    \caption{Binary LOD transition classification architect overview. Similarly, for the multi-class classification the output is an array with the same length as number of quality levels to classify. In this context, the disappearing of a large part of the object (red circle) would be considered a low quality transition.}
    \label{fig:lod_overview}
\end{figure}

\subsection{Binary and multi-class classification}

In this context the binary classification is a method to {\it detect} LOD transitions. An overview can be seen in Figure \ref{fig:lod_overview}. It detects mainly two issues: relative quality difference and transition difference (so called popping). The quality difference can be seen as a noticeable difference in polygon and texture render quality. The transition difference is not necessarily detecting quality difference but how much the structure changes in the transition between two LODs. The binary classification is important as it could be used as a tool to set an "optimal" LOD distance where the transition is not detectable. 

The multi-class classification is trained on different classes of LOD qualities and outputs the perceived quality of a LOD transition. It can be used to automatically detect overall bad transitions or give an indication if the asset has certain angles that the transition is particularly bad.

We evaluate several established computer vision-related architectures, ShuffleNetV2 \cite{ma2018shufflenet}, ResNet \cite{he2016deep}, VGG \cite{simonyan2014very} and MobileNet \cite{sandler2018mobilenetv2} for binary and multi-class classification. As input, we concatenate the image pair in the channel space. For every architecture, we use the pre-trained weights in ImageNet \cite{deng2009imagenet} and duplicate the kernels of the first layer in the channel space to accommodate the image pair input. We remove the last fully connected layer and introduce our own according to the classification problem (binary or multi-class). Finally, we fine-tune the networks with stochastic gradient descent on our custom-made dataset.

To evaluate which of the CNN architectures performed best on the task of detecting LOD reductions, all networks were trained with the same hyperparameters. We select the best performing networks in order to proceed with the hyperparameter search.
We use mean train accuracy, training time and mean validation accuracy to determine performance.

\begin{figure*}[!t]
    \centering
    \includegraphics[width=0.95\textwidth]{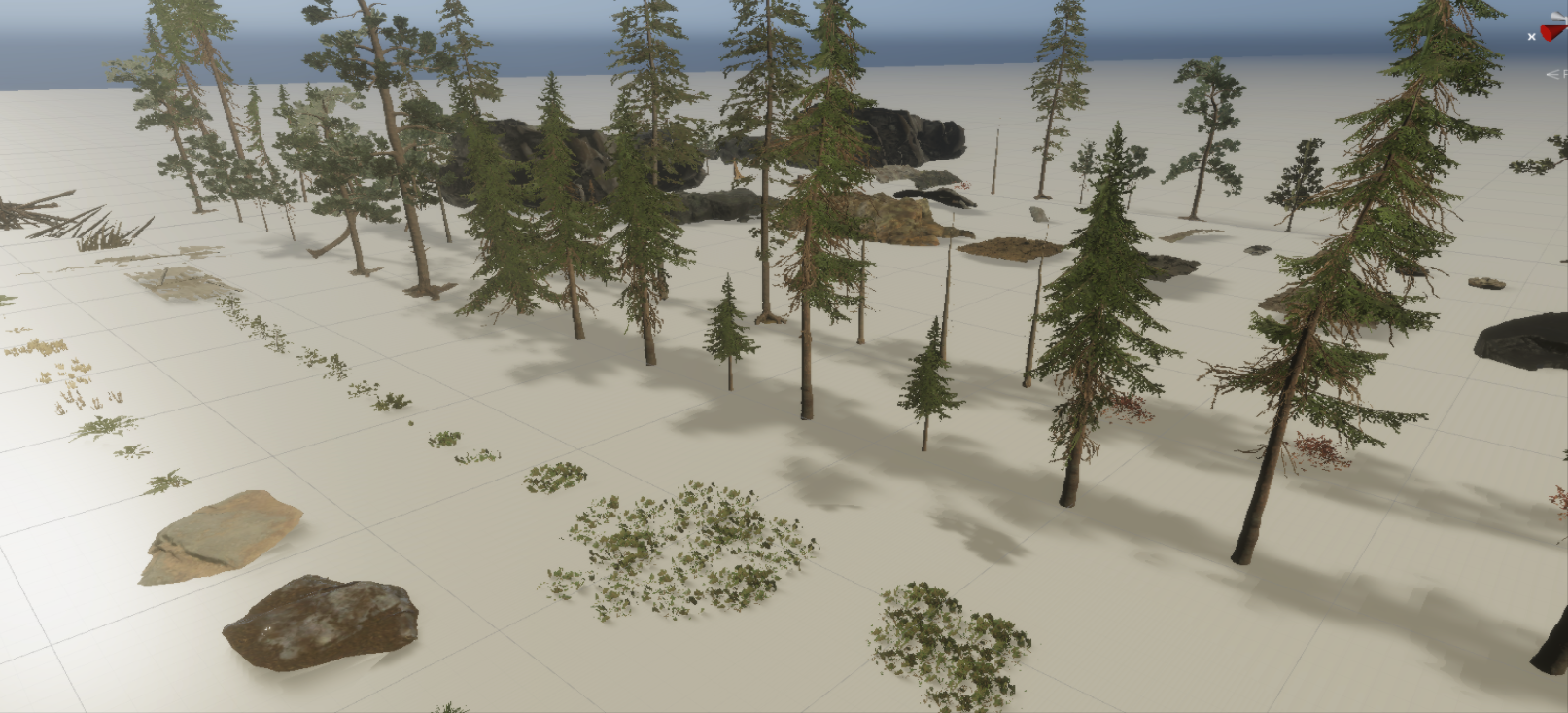}
    \caption{Examples of training data that was used in the training set. Each asset had at least two available different LOD.}
    \label{fig:data_overview}
\end{figure*}

\subsection{Hyperparameters}
After establishing which CNN architecture fits the use-case best and based on the evaluation results, the purpose of this next stage was to fine-tune the model by testing different hyperparameters. The hyperparameters tested are described in this section:

\subsubsection{Learning Rate and Batch Size}
Because the learning rate and batch size are dependent on each other, they were tested together. We used logarithmic exploration and tested a range of values from 10$^-1$ to 10$^-3$, which was further expanded if there was an indication that a lower or higher rate should be tested for better performance.

\subsubsection{Dataset Size}
The dataset size was not used to derive the final deep learning model. Instead, these tests were performed to explore how big the dataset needs to be for the model to learn to detect LOD reductions. The motivation is to determine the amount of new assets needed for the model to learn to detect LOD reductions of those new types.

To explore how big the dataset needed to be, we trained a model using 20, 35, 50 and 65 assets respectively, with 8 rotations and 8 angles per asset resulting in 5120, 8960, 12800 and 16640 pictures respectively. Each bigger dataset was expanded on from the previous dataset, meaning that all the data present in the dataset derived from 20 assets are also present in the dataset derived from 35 assets and so on. The details of the data can be read in Section \ref{sec:data}.


\subsection{Multiclass Classification}\label{sec:multiclass_method}
With the promising results in binary classification, we further explored whether the model's usefulness could be improved using a second method. We apply the idea from the binary classification and instead feed the model multiple versions of LOD reductions. We refer to this method as neural-based multiclass classification or just multiclass classification. The aim of this method is to test whether a deep learning model can be trained to distinguish between LOD reductions of different qualities. Hence, this task is to distinguish between a set (in this case six) of different perceived quality levels.

The data was created as asset pairs, just like for the binary data, where the first picture of the pair is the asset in its full LOD (LOD 0) and the second picture is a variable LOD reduction taken after zooming out a few steps. Unlike the binary data, there are no pairs in which there is not a reduction. Instead, each of these pairs are created six times for an asset (per data augmentation, i.e. with the same angles and rotations) with different amounts of polygon reduction of the second picture in the pair. These six reductions vary in quality from very high to very low, representing six different quality levels where quality level 0 represents LOD 0. The next quality level is 1 which is the least reduced, then 2 which is further reduced and so on, until quality level 6 (which is the lowest quality). Note that quality level is different from LOD level.

The six quality levels were derived using the assets' vertex count. How much the asset is reduced at each level depends on the vertex count of the assets' LOD 0. The higher the vertex count of an asset's LOD 0, the higher the reductions are. The steps were chosen with regards to the number of vertices of the assets present in our datasets. Each reduction amount is relative to the asset's LOD 0 vertex count, i.e. an asset with an original vertices count of 5500 reduced to quality level 4 would be reduced to $5500*0,075$. The LOD reduction steps can be seen in Table \ref{tab:quality_lvls}. The levels were derived using manual experimentation and manual validation. Read more about the process of creating the data for the multiclass classification method in Section \ref{sec:data}. More in-depth description and discussion on the methodology can be found in \cite{tamm22}.

\section{Data Generation}
\label{sec:data}

\begin{figure}[!h]
    \centering
    \includegraphics[width=0.48\textwidth]{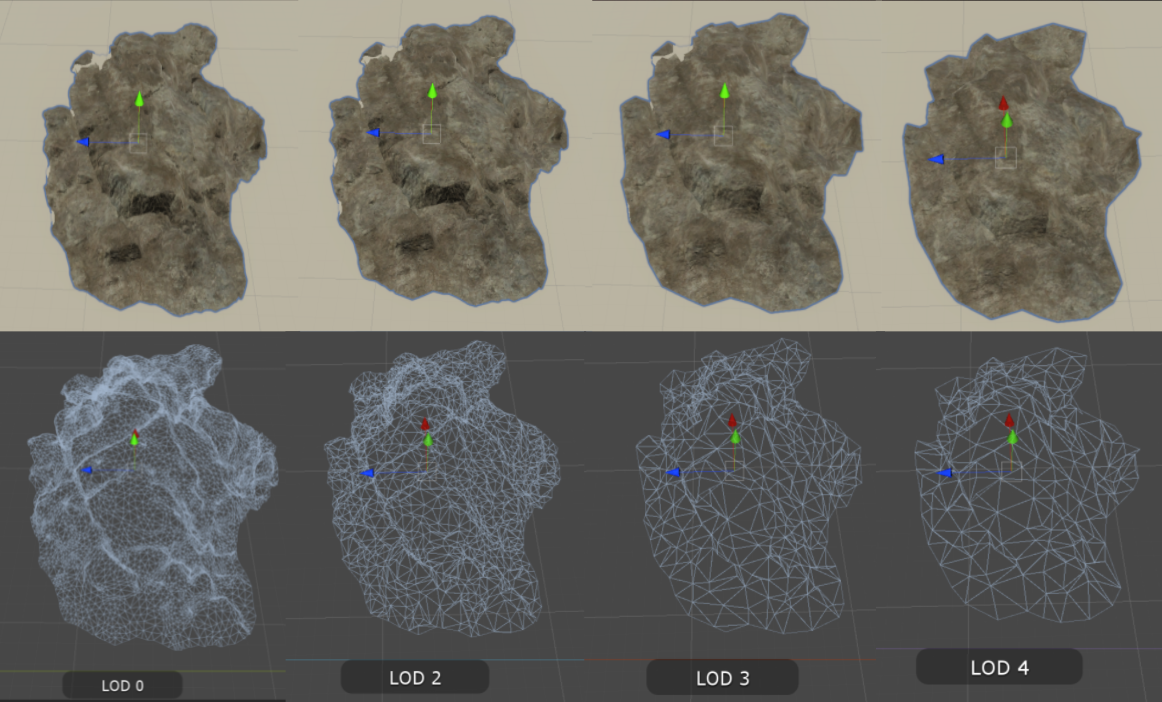}
    \includegraphics[width=0.48\textwidth]{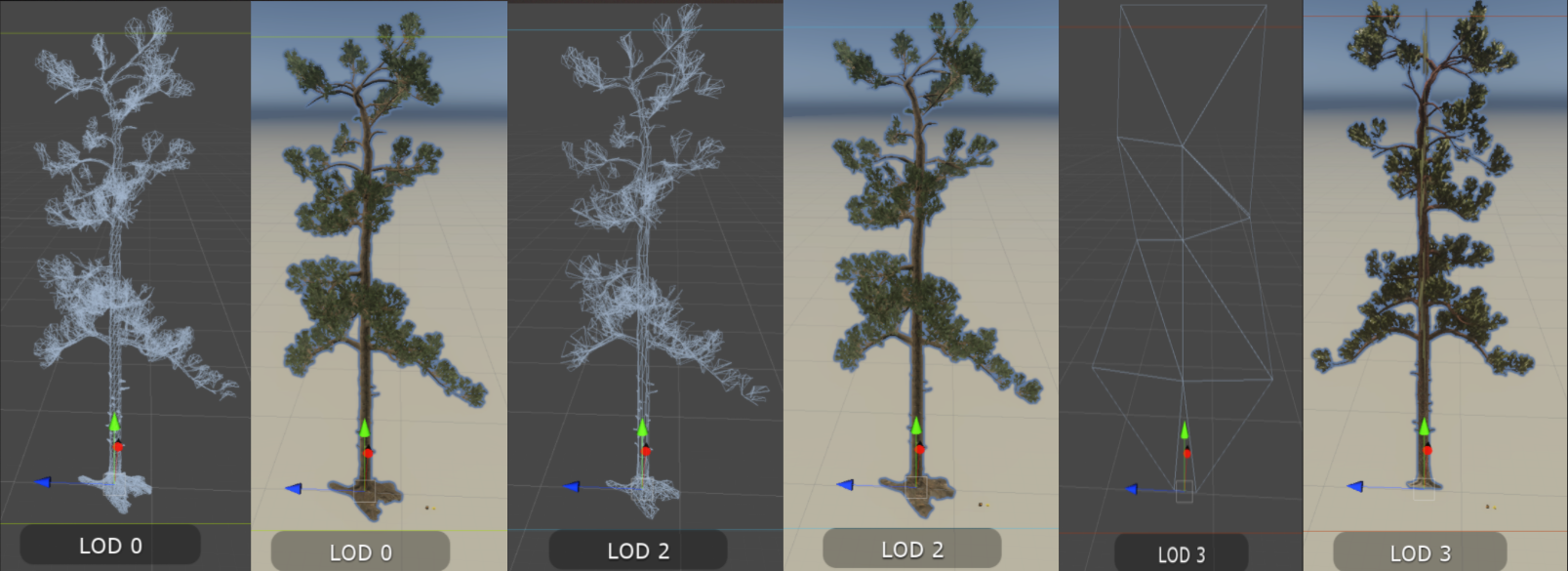}
    \caption{Examples of assets with different LOD.}
    \label{fig:data_samples}
\end{figure}

A dataset with pictures of assets was created for each classification method, see Figure \ref{fig:data_overview} for examples. A detailed image of a sample can be seen in Figure \ref{fig:data_samples}. Each data point consists of a pair of two images, we refer to this as an asset pair. The first image of the pair is the reference image, the highest quality asset or LOD0. In the binary case, the second image of the pair is the same LOD0 or lower, making a balanced dataset of transitions and non-transitions. In the multi-class problem, the second pair is a lower LOD (LOD1-LOD6), and similarly, every LOD has the same appearance ratio.

To create a more robust training dataset and to increase variance in the data, we follow a similar procedure of rendering pictures as \cite{ling2020using, griffin2015evaluating}. Each asset is placed at different distances and LOD levels and is rendered in the game engine with different viewpoints, lighting conditions, backgrounds, etc, resulting in multiple pairs of each asset. Note that between the two pictures in a pair, there are no changes of viewpoints, lighting or backgrounds. The second image of the pair is zoomed out to simulate a possible transition.
The asset pairs are concatenated into a 6-channels picture and labeled with whether there has been a LOD transition or not, see Figure \ref{fig:data_samples} for an example. The model learns to distinguish whether there is a LOD transition or if the asset is just being zoomed out, or of which quality level the transition is of, depending on the classification method used.

\section{Results}
\label{sec:results}
In this section, we evaluate the performance of several CNN architectures for classifying LOD transitions and the ability to detect the perceived quality of LOD transitions.

\subsection{Binary classification}
We present the results from training several CNN architectures to binary classify the existence or not of a LOD transition between a pair of images from the same asset.
First, we compared different CNN architectures.
We train for $10$ epochs and use a 5-fold cross-validation strategy. We believe that this strategy is crucial because there can be assets with different complexity levels. Thus, the way the train-validation split was done would condition the validation score of the model, not showing the real performance of the model. There is no asset intersection in the training and validation images in the data splits.

\begin{figure}[!t]
    \centering
    \includegraphics[width=0.4\textwidth]{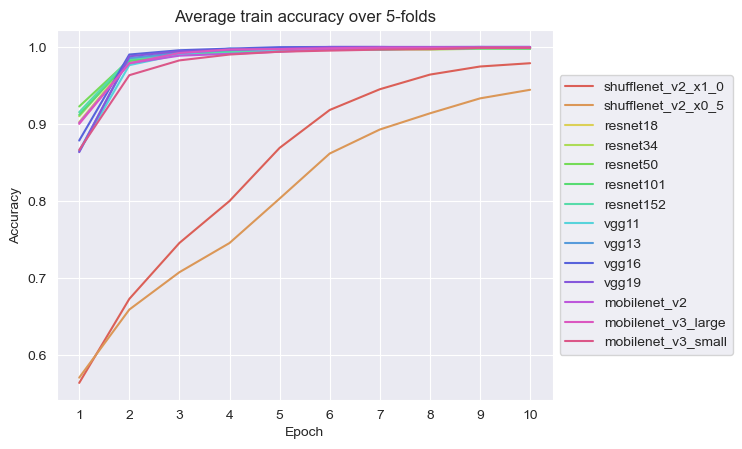}
    \caption{Training accuracy for binary classification.}
    \label{fig:train_accuracy}
\end{figure}

\begin{figure}[!t]
    \centering
    \includegraphics[width=0.4\textwidth]{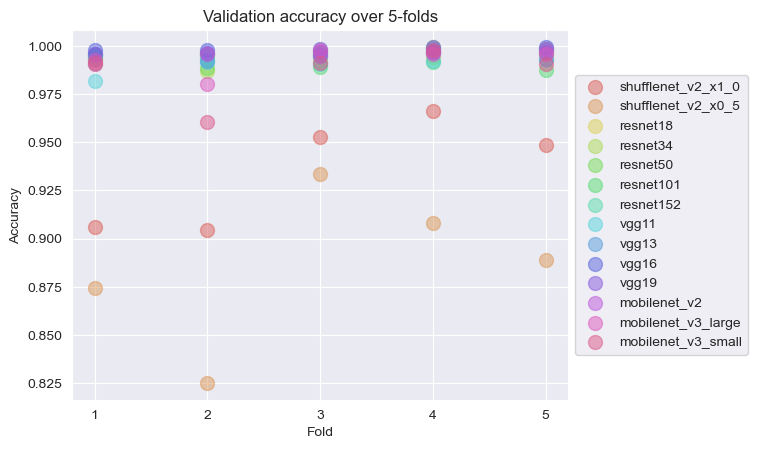}
    \caption{Validation accuracy for binary classification in 10 epochs.}
    \label{fig:validation_accuracy}
\end{figure}

We can observe their training and validation performance in Figure \ref{fig:train_accuracy} and \ref{fig:validation_accuracy} respectively. We notice that, except for ShuffleNetV2, the training converges quickly and does not overfit as the validation accuracy is still high. For a more detailed view of the validation performance, we show the mean accuracy over the folds in Table \ref{tab:all_nets_acc}.
Moving forward for future experiments, we select ResNet18 because of its high validation accuracy and architecture simplicity compared to the other approaches, allowing us faster iterations as the training time is shorter.
We perform a hyperparameter grid search to find the optimal batch size and learning rate shown in Table \ref{tab:lr_b_mean_val_acc}. We conclude that a learning rate of $0.01$ is more suitable while keeping a small batch size of $8$ or $16$. Similarly, we will follow these hyperparameters for future experiments.

\begin{table}[!htb]
\begin{center}
\begin{tabular}{ |c|c| } 
 \hline
  Network & Mean Accuracy \\ 
   \hline\hline
  resnet18 & 0.995 \\ 
  resnet34 & 0.988 \\ 
  resnet50 & 0.994  \\ 
  resnet101 & 0.994  \\ 
  vgg11 & 0.985  \\ 
  vgg13 & 0.995  \\ 
  vgg16 & 0.996  \\ 
  vgg19 & 0.995  \\ 
  mobilenet\_v2 & 0.992  \\ 
  mobilenet\_v3\_large & 0.990  \\ 
  mobilenet\_v3\_small & 0.990  \\ 
 \hline
\end{tabular}
\end{center}\caption{Mean validation accuracies of the networks trained}
\label{tab:all_nets_acc}
\end{table}

\begin{table}
\begin{center}
\begin{tabular}{ |c|c|c|c| } 
 \hline
  Batch Size & 0.1 & 0.01 & 0.001 \\ 
   \hline
  8 & 0.87 & 0.99 & 0.96\\ 
  16 & 0.86 & 0.99 & 0.96 \\ 
  32 & 0.90 & 0.98 & 0.95\\ 
  64 & 0.88 & 0.98 & 0.94\\ 
  128 & 0.91 & 0.97 & 0.90\\ 
 \hline
\end{tabular}
\end{center}\caption{Mean validation accuracy. Hyperparameter grid search for batch size and learning rate.}
\label{tab:lr_b_mean_val_acc}
\end{table}

\begin{table}[!t]
\scalebox{0.7}{
\begin{tabular}{l|llllll}
Dataset & Large pines & Small pines & Stump & Tree dead & Wood log & Wood log 2 \\ \hline
20 stones assets        & 0.91        & 0.93        & 0.94  & 0.78      & 0.84     & 0.50       \\
35 stones assets        & 0.70        & 0.91        & 0.62  & 0.75      & 0.62     & 0.97       \\
50 stones assets        & 0.70        & 0.85        & 0.59  & 0.59      & 0.50     & 0.88       \\
65 stones assets        & 0.77        & 0.89        & 0.56  & 0.59      & 0.59     & 0.78       \\
Bush \& rock  assets    & 1.00       & 1.00  & 0.81      & 0.88     & 0.72     & 1.00 \\
\end{tabular}
}
\caption{Test accuracy results of different types of assets with binary classification. Training parameters: ResNet18 batch size of 8, 10 epochs, a  learning rate of 0.01 and SGD optimizer.}
\label{tab:test_acc_results_diff_datasets}
\end{table}

We can find a high variability of assets (i.e. stones, trees, bushes, etc.) that can increase throughout the lifetime of a project in game production.
We evaluate the performance of our selected CNN architecture, ResNet18, and hyperparameters, batch size of $8$ and learning rate of $0.01$ trained in several datasets with various sizes. We want to explore the necessary amount of required data in order to train our model and to generalize over unseen asset types. We show in Table \ref{tab:test_acc_results_diff_datasets} the accuracy results for various test datasets (i.e, trees, stumps, wood logs) that are unseen by the model in the training data. We also evaluate our model with several training datasets varying on asset count number (do not confuse with rendered image pairs) consisting of rocks. We can observe that the model already performs adequately for unseen asset types with a low number of asset rocks. Adding more assets of the same kind does not positively impact the accuracy. In addition and as expected, we notice that including an asset type closer in perceived appearance to the test data improves the accuracy overall to detect a LOD transition. In this case, bushes are more similar to trees than rocks are. To sum up, we show that the model can generalize to unseen asset types trained with a reduced number of assets of a different type and the accuracy improves by including an asset of similar perceived appearance.

\subsection{Multi-class classification - Quality}
We evaluate the previous ResNet18 architecture to detect the perceived quality of LOD transitions. We set the problem as a multi-class classification task where the model predicts the quality of the LOD transition by distinguishing between different LOD reductions. We create a dataset with $6$ different LOD reductions, and as can be seen in Table \ref{tab:quality_lvls}, each reduction step is relative to the asset's original vertices count in our binary dataset. The reductions are procedurally made using Fast-Quadric-Mesh-Simplification \cite{fastquadratic}. 
The input to the model is a pair of images, where the first of the pair is the reference high-quality LOD0 image and the second of the pair is one of the LOD reduction rendered image.

\begin{table}
\begin{center}
\begin{tabular}{|c|c c c c|} 
 \hline
  & \multicolumn{4}{ c |}{Vertices Count}  \\ [0.5ex] 
 \hline
 Quality Level & $<100$ & $<300$ & $<500$ & $>=500$ \\ 
 \hline
 1 & 0,98 & 0,9 & 0,9 & 0,5 \\ 
 \hline
 2 & 0,92 & 0,81 & 0,72 & 0,25 \\
 \hline
 3 & 0,87 & 0,73 & 0,5 & 0,13 \\
 \hline
 4 & 0,81 & 0,58 & 0,3 & 0,075 \\
 \hline
 5 & 0,76 & 0,47 & 0,15 & 0,045 \\
 \hline
 6 & 0,7 & 0,37 & 0,075 & 0,027 \\ [1ex] 
 \hline
\end{tabular}
\end{center}\caption{The quality reduction used for each quality level based on the vertices count of the asset.}
\label{tab:quality_lvls}
\end{table}

\begin{figure}[!t]
    \centering
\includegraphics[width=0.4\textwidth]{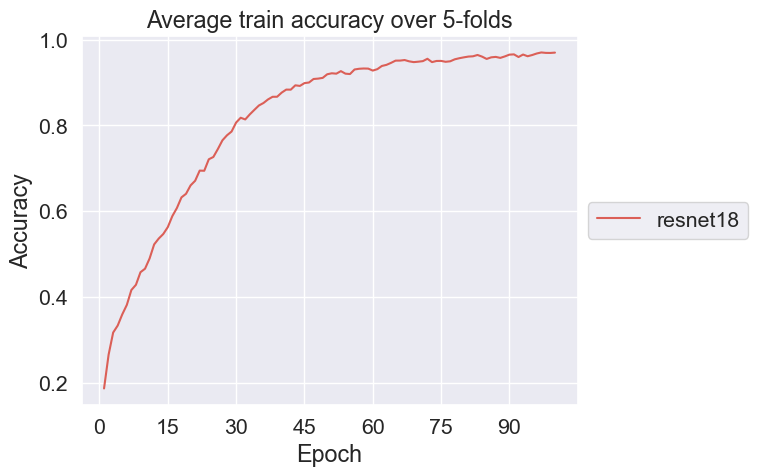}
    \includegraphics[width=0.4\textwidth]{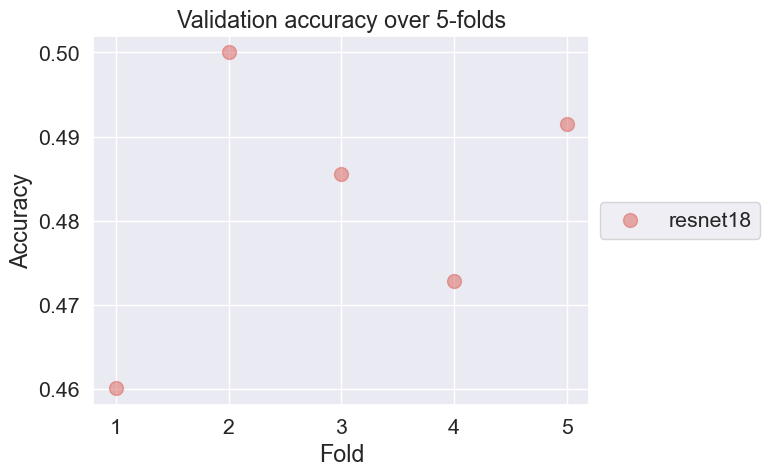}
    \caption{Train and validation accuracy for multiclass.}
    \label{fig:accuracy_multiclass}
\end{figure}

\begin{figure}[!t]
    \centering
    \includegraphics[width=0.4\textwidth]{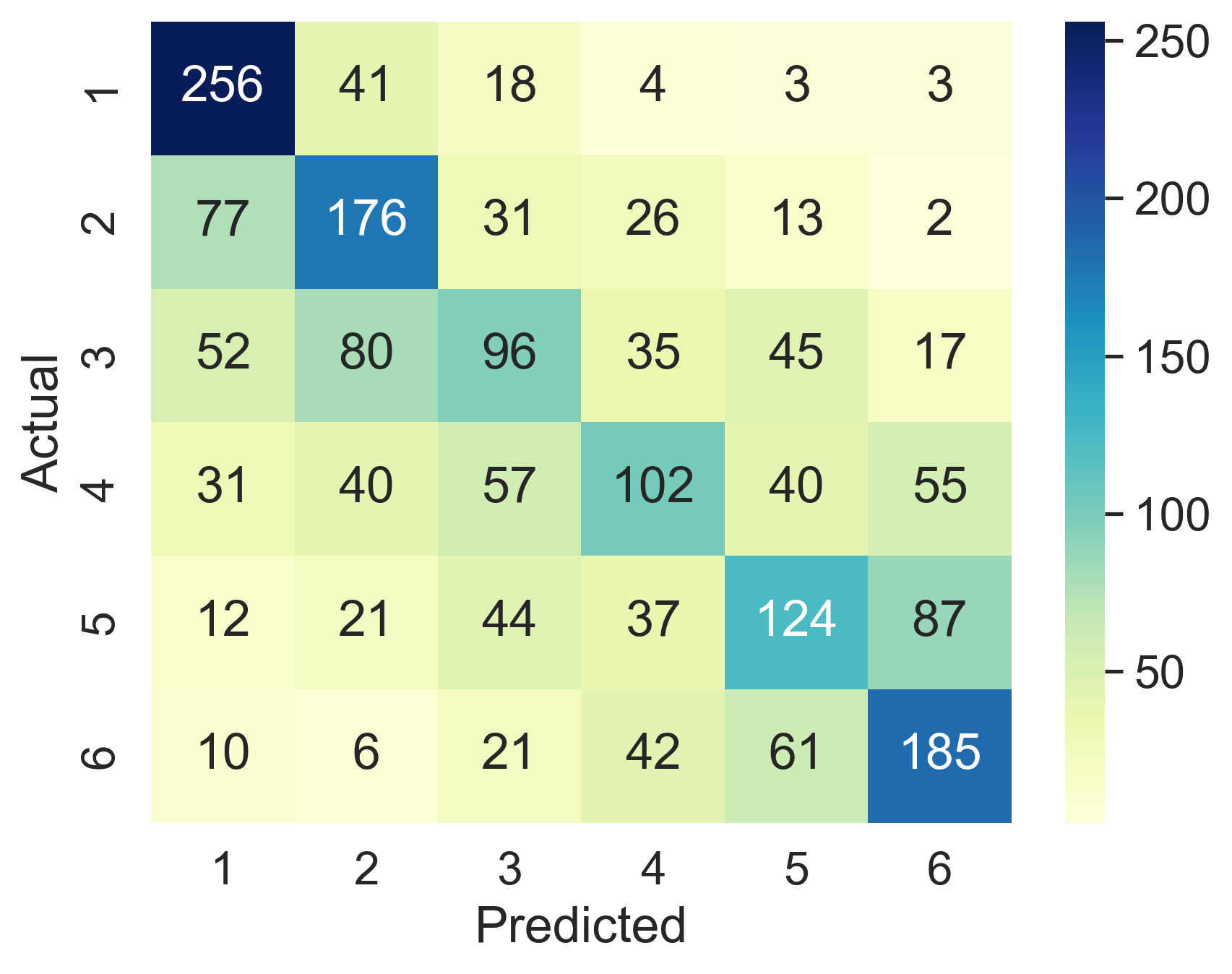}
    \caption{Confusion matrix for validation set}
    \label{fig:multiclass_confusion_matrix_val}
\end{figure}

We train the CNN architecture with the same hyperparameters as the binary classification problem for $100$ epochs instead of $10$ as the dataset is smaller. We can observe the training and validation performance in Figure \ref{fig:accuracy_multiclass}. We notice that the overall mean accuracy is not really high. We then focus on the validation confusion matrix in Figure \ref{fig:multiclass_confusion_matrix_val} to see how the model performs class-wise. The model tends to confuse the classification of adjacent quality classes, such as confusing class 2 for a class 1 or 3. We can also see that this misclassification is gradual in the sense that class 1 is misclassified as class 2 more times than it is misclassified as class 3. Therefore, the overall accuracy rate is rather low. However, these results are acceptable as similar results would be expected if a human would attempt to classify the transitions since the difference between adjacent classes is relatively small.

To compare with the binary classification, we merged the classes 1-3 and 4-6 to be able to compare quantitatively, assuming the respective ground corresponds to bad vs. good transition. The accuracy is 80\%, lower than in the case of binary classification. However, we see that most of the misclassifications happen in the neighbor classes 3-4. Excluding these classes (they are somewhat ambiguous as just polygon count ratio procedurally classifies the assets’ quality level), the accuracy is 94.8\%, closer to that of the binary case.

\begin{figure}[!t]
    \centering
    \includegraphics[width=0.4\textwidth]{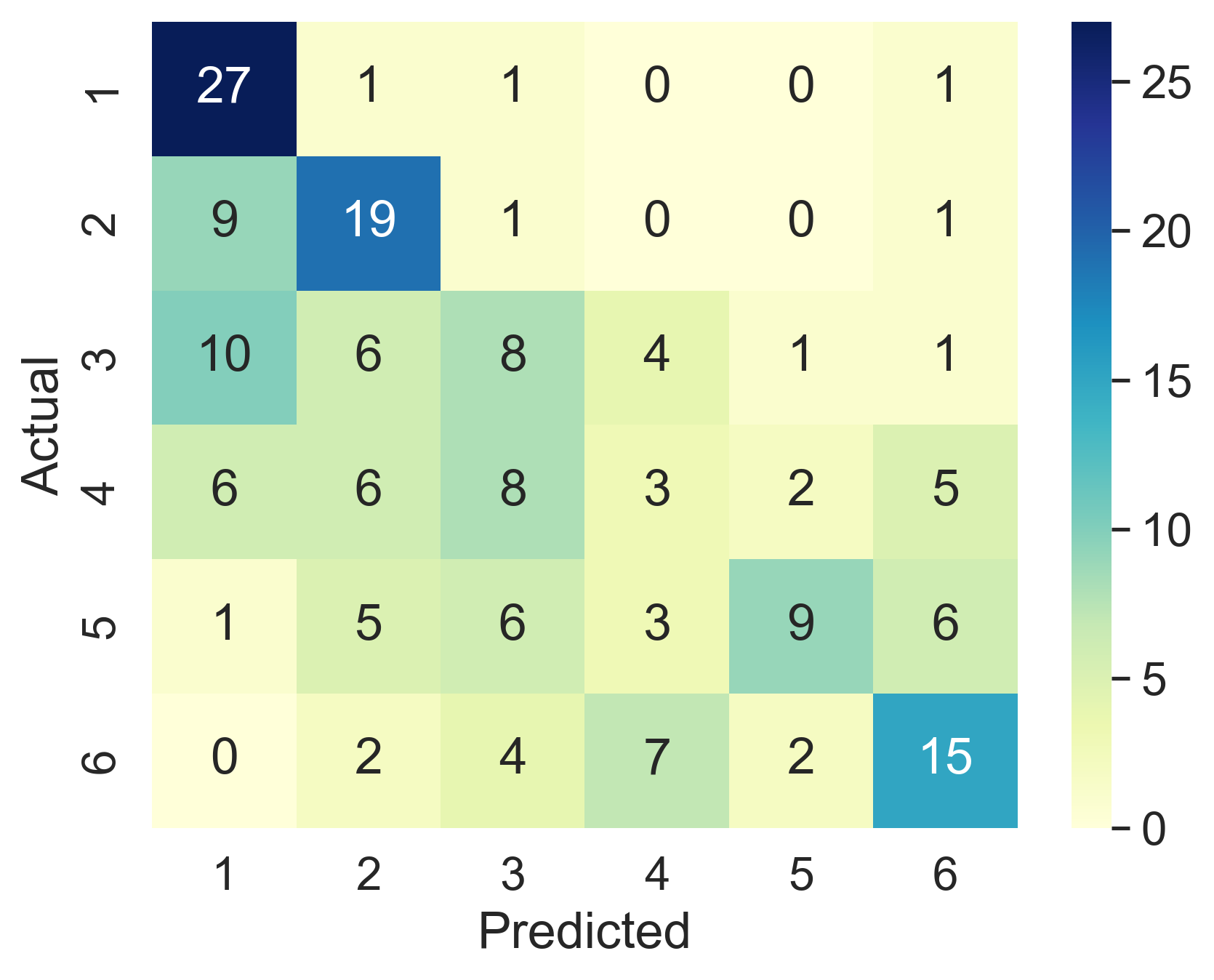}
    \caption{Confusion matrix for final test set}
    \label{fig:multiclass_confusion_matrix_test}
\end{figure}

Additionally, a final evaluation was done on a test dataset with assets never seen before by the model. The results can be seen in  Figure \ref{fig:multiclass_confusion_matrix_test}. The test data's confusion matrix shows similar behavior to the validation data one. Merging the classes the same way as the validation results shows a 74\% accuracy.

\section{Conclusion and Future Work}
\label{sec:conclusion}

We have shown that deep convolutional neural networks is a promising approach for automated LOD reduction validation. We studied mainly two approaches, binary classification: we train a model to detect a transition between two LODs, and multi-class: train a model to estimate the perceived quality of a LOD transition. For example, the former can be used to optimize game environment scenes to the lowest polygon count detectable (by humans) levels. The latter can be used as an assistant tool to verify the quality of a LOD from all different angles, lighting conditions, etc. automatically. Our findings show that indeed this can be done and we can give recommendations on models to use, input format, training data, etc. Our conclusion is that it is always better to set up a dataset that is similar to the data at hand, but it does not have to be identical as even fitting a moderate big dataset, we can see that it can generalize over other type of unseen asset types. Our models can be used to verify large quantities of assets formed by several LODs. The procedure would require rendering pair of images formed by the reference quality and the reduced LOD asset from different viewpoints and lighting conditions, as stated in Section \ref{sec:data}. The quality multi-class classification network estimates the quality of the transitions. It will be possible to identify which LOD assets do not meet the desired standards. Similarly, this same approach could be used to compare different automatic LOD reduction tools.

As previously mentioned, we have identified at least two interesting future work with this approach. The first area of research is scene optimization, where we envision that this approach could be used as a method to optimize the distances that the LOD switched to lower resolution. This could lead to a method to optimize full scenes where the LOD can be set to occur as early as possible without affecting the perceived quality. Secondly, we envision this approach can be used as a discriminator network for an adversarial approach and to help improve the training of LOD reduction tools, or to help with selecting the right LOD reduction tool that fits the specific need of the project.

\section{Acknowledgements}

The authors would like to thank Gianvito Serra (DICE EA), Jonathan Greenberg (SEED EA) and Alan Wolfe (SEED EA) for valuable feedback and discussions. 

\bibliography{refs}

\begin{thebibliography}{10}
\providecommand{\url}[1]{#1}
\csname url@samestyle\endcsname
\providecommand{\newblock}{\relax}
\providecommand{\bibinfo}[2]{#2}
\providecommand{\BIBentrySTDinterwordspacing}{\spaceskip=0pt\relax}
\providecommand{\BIBentryALTinterwordstretchfactor}{4}
\providecommand{\BIBentryALTinterwordspacing}{\spaceskip=\fontdimen2\font plus
\BIBentryALTinterwordstretchfactor\fontdimen3\font minus
  \fontdimen4\font\relax}
\providecommand{\BIBforeignlanguage}[2]{{%
\expandafter\ifx\csname l@#1\endcsname\relax
\typeout{** WARNING: IEEEtran.bst: No hyphenation pattern has been}%
\typeout{** loaded for the language `#1'. Using the pattern for}%
\typeout{** the default language instead.}%
\else
\language=\csname l@#1\endcsname
\fi
#2}}
\providecommand{\BIBdecl}{\relax}
\BIBdecl

\bibitem{turk2005stanford}
G.~Turk and M.~Levoy, ``The stanford 3d scanning repository,'' \emph{Stanford
  University Computer Graphics Laboratory http://graphics. stanford.
  edu/data/3Dscanrep}, vol.~1, p.~5, 2005.

\bibitem{lod3dgraphics}
D.~Luebke, M.~Reddy, J.~D.Cohen, A.~Varshney, B.~Watson, and R.~Huebner,
  \emph{Level of Detail For 3D Graphics}.\hskip 1em plus 0.5em minus
  0.4em\relax Morgan Kaufmann Publishers, 2003.

\bibitem{schwarz2006gpu}
M.~Schwarz, M.~Staginski, and M.~Stamminger, ``Gpu-based rendering of pn
  triangle meshes with adaptive tessellation,'' in \emph{Proceedings of Vision,
  Modeling, and Visualization}.\hskip 1em plus 0.5em minus 0.4em\relax
  Akademische Verlagsgesellschaft Aka GmbH, Berlin, 2006, pp. 161--168.

\bibitem{bergdahl2020augmenting}
J.~Bergdahl, C.~Gordillo, K.~Tollmar, and L.~Gissl{\'e}n, ``Augmenting
  automated game testing with deep reinforcement learning,'' in \emph{2020 IEEE
  Conference on Games (CoG)}.\hskip 1em plus 0.5em minus 0.4em\relax IEEE,
  2020, pp. 600--603.

\bibitem{holmgaard2018automated}
C.~Holmg{\aa}rd, M.~C. Green, A.~Liapis, and J.~Togelius, ``Automated
  playtesting with procedural personas through mcts with evolved heuristics,''
  \emph{IEEE Transactions on Games}, vol.~11, no.~4, pp. 352--362, 2018.

\bibitem{gordillo2021improving}
C.~Gordillo, J.~Bergdahl, K.~Tollmar, and L.~Gissl{\'e}n, ``Improving
  playtesting coverage via curiosity driven reinforcement learning agents,''
  \emph{arXiv preprint arXiv:2103.13798}, 2021.

\bibitem{chandler2013seven}
D.~M. Chandler, ``Seven challenges in image quality assessment: past, present,
  and future research,'' \emph{International Scholarly Research Notices}, vol.
  2013, 2013.

\bibitem{andersson2020flip}
P.~Andersson, J.~Nilsson, T.~Akenine-M{\"o}ller, M.~Oskarsson,
  K.~{\AA}str{\"o}m, and M.~D. Fairchild, ``Flip: A difference evaluator for
  alternating images.'' \emph{Proc. ACM Comput. Graph. Interact. Tech.},
  vol.~3, no.~2, pp. 15--1, 2020.

\bibitem{wolski2018dataset}
K.~Wolski, D.~Giunchi, N.~Ye, P.~Didyk, K.~Myszkowski, R.~Mantiuk, H.-P.
  Seidel, A.~Steed, and R.~K. Mantiuk, ``Dataset and metrics for predicting
  local visible differences,'' \emph{ACM Transactions on Graphics (TOG)},
  vol.~37, no.~5, pp. 1--14, 2018.

\bibitem{amirshahi2016image}
S.~A. Amirshahi, M.~Pedersen, and S.~X. Yu, ``Image quality assessment by
  comparing cnn features between images,'' \emph{Journal of Imaging Science and
  Technology}, vol.~60, no.~6, pp. 60\,410--1, 2016.

\bibitem{bosse2017deep}
S.~Bosse, D.~Maniry, K.-R. M{\"u}ller, T.~Wiegand, and W.~Samek, ``Deep neural
  networks for no-reference and full-reference image quality assessment,''
  \emph{IEEE Transactions on image processing}, vol.~27, no.~1, pp. 206--219,
  2017.

\bibitem{wang2003multiscale}
Z.~Wang, E.~P. Simoncelli, and A.~C. Bovik, ``Multiscale structural similarity
  for image quality assessment,'' in \emph{The Thrity-Seventh Asilomar
  Conference on Signals, Systems \& Computers, 2003}, vol.~2.\hskip 1em plus
  0.5em minus 0.4em\relax Ieee, 2003, pp. 1398--1402.

\bibitem{zhang2014vsi}
L.~Zhang, Y.~Shen, and H.~Li, ``Vsi: A visual saliency-induced index for
  perceptual image quality assessment,'' \emph{IEEE Transactions on Image
  processing}, vol.~23, no.~10, pp. 4270--4281, 2014.

\bibitem{mantiuk2011hdr}
R.~Mantiuk, K.~J. Kim, A.~G. Rempel, and W.~Heidrich, ``Hdr-vdp-2: A calibrated
  visual metric for visibility and quality predictions in all luminance
  conditions,'' \emph{ACM Transactions on graphics (TOG)}, vol.~30, no.~4, pp.
  1--14, 2011.

\bibitem{wang2004image}
Z.~Wang, A.~C. Bovik, H.~R. Sheikh, and E.~P. Simoncelli, ``Image quality
  assessment: from error visibility to structural similarity,'' \emph{IEEE
  transactions on image processing}, vol.~13, no.~4, pp. 600--612, 2004.

\bibitem{nilsson2020understanding}
J.~Nilsson and T.~Akenine-M{\"o}ller, ``Understanding ssim,'' \emph{arXiv
  preprint arXiv:2006.13846}, 2020.

\bibitem{wolski2019selecting}
K.~Wolski, D.~Giunchi, S.-i. Kinuwaki, P.~Didyk, K.~Myszkowski, A.~Steed, and
  R.~K. Mantiuk, ``Selecting texture resolution using a task-specific
  visibility metric,'' in \emph{Computer Graphics Forum}, vol.~38, no.~7.\hskip
  1em plus 0.5em minus 0.4em\relax Wiley Online Library, 2019, pp. 685--696.

\bibitem{krizhevsky2012imagenet}
A.~Krizhevsky, I.~Sutskever, and G.~E. Hinton, ``Imagenet classification with
  deep convolutional neural networks,'' \emph{Advances in neural information
  processing systems}, vol.~25, 2012.

\bibitem{prashnani2018pieapp}
E.~Prashnani, H.~Cai, Y.~Mostofi, and P.~Sen, ``Pieapp: Perceptual image-error
  assessment through pairwise preference,'' in \emph{Proceedings of the IEEE
  Conference on Computer Vision and Pattern Recognition}, 2018, pp. 1808--1817.

\bibitem{bradley1952rank}
R.~A. Bradley and M.~E. Terry, ``Rank analysis of incomplete block designs: I.
  the method of paired comparisons,'' \emph{Biometrika}, vol.~39, no. 3/4, pp.
  324--345, 1952.

\bibitem{zhang2018unreasonable}
R.~Zhang, P.~Isola, A.~A. Efros, E.~Shechtman, and O.~Wang, ``The unreasonable
  effectiveness of deep features as a perceptual metric,'' in \emph{Proceedings
  of the IEEE conference on computer vision and pattern recognition}, 2018, pp.
  586--595.

\bibitem{iandola2016squeezenet}
F.~N. Iandola, S.~Han, M.~W. Moskewicz, K.~Ashraf, W.~J. Dally, and K.~Keutzer,
  ``Squeezenet: Alexnet-level accuracy with 50x fewer parameters and< 0.5 mb
  model size,'' \emph{arXiv preprint arXiv:1602.07360}, 2016.

\bibitem{simonyan2014very}
K.~Simonyan and A.~Zisserman, ``Very deep convolutional networks for
  large-scale image recognition,'' \emph{arXiv preprint arXiv:1409.1556}, 2014.

\bibitem{lodTexture3DMaps}
\BIBentryALTinterwordspacing
M.~U. Hiroyuki~Inatsuka and M.~Okuda, ``Level of detail control for texture on
  3d maps,'' 2005. [Online]. Available:
  \url{https://ieeexplore.ieee.org/document/1524288}
\BIBentrySTDinterwordspacing

\bibitem{guo2016subjective}
J.~Guo, V.~Vidal, I.~Cheng, A.~Basu, A.~Baskurt, and G.~Lavoue, ``Subjective
  and objective visual quality assessment of textured 3d meshes,'' \emph{ACM
  Transactions on Applied Perception (TAP)}, vol.~14, no.~2, pp. 1--20, 2016.

\bibitem{pan2005quality}
Y.~Pan, I.~Cheng, and A.~Basu, ``Quality metric for approximating subjective
  evaluation of 3-d objects,'' \emph{IEEE Transactions on Multimedia}, vol.~7,
  no.~2, pp. 269--279, 2005.

\bibitem{hasselgren2021appearance}
J.~Hasselgren, J.~Munkberg, J.~Lehtinen, M.~Aittala, and S.~Laine,
  ``Appearance-driven automatic 3d model simplification,'' \emph{arXiv preprint
  arXiv:2104.03989}, 2021.

\bibitem{ling2020using}
C.~Ling, K.~Tollmar, and L.~Gissl{\'e}n, ``Using deep convolutional neural
  networks to detect rendered glitches in video games,'' in \emph{Proceedings
  of the AAAI Conference on Artificial Intelligence and Interactive Digital
  Entertainment}, vol.~16, no.~1, 2020, pp. 66--73.

\bibitem{ma2018shufflenet}
N.~Ma, X.~Zhang, H.-T. Zheng, and J.~Sun, ``Shufflenet v2: Practical guidelines
  for efficient cnn architecture design,'' in \emph{Proceedings of the European
  conference on computer vision (ECCV)}, 2018, pp. 116--131.

\bibitem{he2016deep}
K.~He, X.~Zhang, S.~Ren, and J.~Sun, ``Deep residual learning for image
  recognition,'' in \emph{Proceedings of the IEEE conference on computer vision
  and pattern recognition}, 2016, pp. 770--778.

\bibitem{sandler2018mobilenetv2}
M.~Sandler, A.~Howard, M.~Zhu, A.~Zhmoginov, and L.-C. Chen, ``Mobilenetv2:
  Inverted residuals and linear bottlenecks,'' in \emph{Proceedings of the IEEE
  conference on computer vision and pattern recognition}, 2018, pp. 4510--4520.

\bibitem{deng2009imagenet}
J.~Deng, W.~Dong, R.~Socher, L.-J. Li, K.~Li, and L.~Fei-Fei, ``Imagenet: A
  large-scale hierarchical image database,'' in \emph{2009 IEEE conference on
  computer vision and pattern recognition}.\hskip 1em plus 0.5em minus
  0.4em\relax Ieee, 2009, pp. 248--255.

\bibitem{tamm22}
M.~Tamm and S.~Olivia, ``Automatic validation of lod reductions through deep
  learning,'' 2022.

\bibitem{griffin2015evaluating}
W.~Griffin and M.~Olano, ``Evaluating texture compression masking effects using
  objective image quality assessment metrics,'' \emph{IEEE Transactions on
  Visualization and Computer Graphics}, vol.~21, no.~8, pp. 970--979, 2015.

\bibitem{fastquadratic}
S.~Forstmann, ``Fast-quadric-mesh-simplification,''
  \url{https://github.com/sp4cerat/Fast-Quadric-Mesh-Simplification}, 2016.

\end{thebibliography}
\bibliographystyle{IEEEtran}

\end{document}